\documentclass[journal]{IEEEtran}

\usepackage{amsmath}
\usepackage{algpseudocode,algorithm,algorithmicx}
\usepackage{dsfont}
\usepackage{hyperref}
\usepackage{graphicx}
\usepackage{amssymb}
\usepackage{amsmath}
\usepackage{mathtools}
\usepackage{cite}
\usepackage{amsthm}
\usepackage{epsfig}
\usepackage{stfloats}
\usepackage{subfigure}
\usepackage{psfrag}
\usepackage[mathscr]{euscript}
\usepackage{acronym}
\usepackage{booktabs}
\usepackage{epstopdf}
\usepackage{pstool}
\usepackage{multirow}
\usepackage{verbatim}
\usepackage{enumitem}

\definecolor{blue2}{rgb}{1,0.2,0.2}
\ifCLASSINFOpdf
\else
\fi
\begin{document}
\title{6G: from Densification to Diversification}

\author{Hyunsoo Kim,
        Taehyung Kim,
        Hyejin Kim,
        Insik Jung,
        Hakkeon Lee, 
        Hyunmin Seo,~\IEEEmembership{Student~Member,~IEEE,} \\
        and Daesik Hong,~\IEEEmembership{Fellow,~IEEE}% <-this % stops a space
%\thanks{M. Shell was with the Department
%of Electrical and Computer Engineering, Georgia Institute of Technology, Atlanta,
} 

\maketitle

\begin{abstract}
The 5G system has finally begun commercialization, and now is the time to start discussing the road map for the 6G system. While the 5G system was designed with a focus on discovering new service types for high speed, low-latency, and massive connective services, the evolution of the network interface for 6G should be considered with an eye toward supporting these complicated communication environments. As machine-driven data traffic continues to increase exponentially, 6G must be able to support a series of connection methods that did not previously exist. In departure from base-station-oriented cell densification, network diversification is necessary if we are to satisfy the comprehensive requirements of end terminals for diverse applications. In this article, we predict what will drive 6G and look at what key requirements should be considered in 6G. We then diversify four types of network architectures according to link characteristics, communication ranges, and target services. The four types of networks play complementary roles while at the same time collaborating across the entire 6G network. Lastly, we call attention to key technologies and challenges in the air, network, and assistive technologies that will have to be addressed when designing the 6G system.

\end{abstract}

\begin{IEEEkeywords}
6G, Network Diversification, Super Proximus Network, Event-Centric Self-Organizing Network, Wide-Area Control Network, Integrated Sky Network.
\end{IEEEkeywords}

% \fi
\IEEEpeerreviewmaketitle

\section{Introduction}
\IEEEPARstart{W}{ireless} mobile communication systems have evolved to complement previous generations, and these advances have been accompanied by changes in killer applications. From the first generation (1G) to 4G, voice call services were quickly converted into data-centric services \cite{5G}. Recently, the Digital Transformation has led to the emergence of disruptive services requiring massive access, high reliability, and low latency, as well as a high data rate. In this flow, 5G defines three types of services: enhanced mobile broadband (eMBB), ultra-reliable low-latency communications (URLLC), and massive machine type communications (mMTC).

Recently, lots of research groups have been aggressively conducting research to lay the foundation for 6G. The International Telecommunication Union (ITU) established the Focus Group Technologies for Network 2030 \cite{ITU}. As part of the 6G flagship research program, the University of Oulu in Finland  has published a white paper for 6G ubiquitous wireless intelligence \cite{Finland}. Samsung Electronics in South Korea is accelerating research by establishing 6G research center \cite{Samsung}.

In particular, machine-made traffic produced by intelligent devices will create new needs in the 5G network and beyond. Unlike multimedia-oriented smartphone traffic, machine-made traffic requires various types of communication links and requirements depending on the characteristics of the target application. To cope with the machine society, the next generation communication system must be able to support ultra-fast transmission for big data processing and have an expandable structure. In addition, the system must provide infinite coverage that can be connected anytime, anywhere.

This is the right time to start designing for `\textit{flexibility-to-any-service},' which means networks intelligently associate users or machines to the most appropriate type of network without the users’ knowledge. However, despite a variety of changes in communication environments, the base-station(BS)-oriented network structure has persisted until the 5G system. Even in 6G environments where heterogeneous requirements have to be satisfied, is this network structure an efficient approach? Before answering that question, let's sketch out how the upcoming 6G communication environment is likely to change. 

\section{What Drives 6G?} 
\vspace{2mm}

\subsection{Future Environment from 3 Perspectives}
\vspace{2mm}
\begin{itemize}[leftmargin=*]
\item \textbf{Human-to-Human Communications: New Media \& UI}
\end{itemize}

The development of mobile telecommunications has completely changed the form of media people use to communicate. Recently, the need for non-face-to-face communication media in industry and education has emerged due to the influence of COVID-19. In 6G, the five senses-based hologram service will emerge, which requires vast amounts of data transmission and ultra-low latency to provide realistic three-dimensional (3D) video with tactile information.

The emergence of multi-dimension media will also make a big difference in the user interface (UI). There will be radio interfaces with powerful computing performance nearby that can handle huge amounts of data, so it will no longer be necessary for people to carry a high-end device for computational tasks. A communication system that actively utilizes cloud computing entails frequent information exchange with surrounding access points and cloud computers, and the UI will have to be modified accordingly. Along this line, smart-lens and glasses can be widely used as well as display-only devices. Foldable and wearable computers such as paper-type devices and implanted sensors will be the mainstream of the 6G UI.

\vspace{2mm}

\begin{itemize}[leftmargin=*]
\item \textbf{Human-to-Machine Communications: Smart Human}
\end{itemize}

With the industrialization of the information and communications technologies (ICT) in the non-ICT field brought about by the Digital Transformation, the digital big blur phenomenon (blurring of the boundary between real life and the digital world) will be accelerated. In particular, human and computer collaboration (HCC) will be popularized. This is an innovation in the human cognitive behavioral system that involves computer functions in a series of processes in which humans perceive and make judgments. For example, HCC can be applied to diagnose diseases by doing computer analysis of 16K ultra-high resolution images and various biological signals that humans cannot analyze. HCC will require not only the ability to receive massive amounts of data from a variety of sources, but also tremendous computing power and information visualization technologies to show the processed information.
\vspace{2mm}

\begin{itemize}[leftmargin=*]
\item \textbf{Machine-to-Machine Communications: RPA Society}
\end{itemize}

Beyond the passive information collection of IoT systems considered in 5G, a robotic process automation (RPA) society will be formed where machines can communicate and produce alone without human intervention. Connected machine systems including industrial robots and autonomous vehicles will expand into a variety of fields, including medicine, education, delivery, retail, construction, and disaster recovery. 

In addition, as new transportation types such as flying vehicles, hyperloops, and underwater highways emerge, new demand will arise in areas where vehicle-to-everything (V2X) network infrastructures were not required before. Since a high speed vehicle in harsh environments transmits and computes large amounts of data and takes action based on the results and interpretations, the communication system is highly relevant to safety. Therefore, 6G networks will have to be able to satisfy more stringent reliability and resilience requirements via infinite coverage.

\subsection{Requirements for 6G}
Before we can consider the technical description of the 6G network, it is necessary to identify the key requirements for the 6G scenarios described above. It should be kept in mind that the following requirements cannot all be met at the same time, but can be achieved through the collaboration of diversified 6G networks as described in Section III below.
\vspace{1mm}

\begin{itemize}[leftmargin=*]
\item \textbf{Extreme High Speed Data Rate}
\end{itemize}

Improving the data rate is a steady challenge from 1G to 5G. According to the ITU, a hologram of $77 \times 20$ inches of human size, colors, full parallax, and $30$ fps will require a speed of $4.62$ Tbps. In addition, as the need for big data for machines to analyze becomes more important, it will be necessary to transfer data in ever larger quantities and volumes. Maximum $1$ Tbps machine experienced data rate and $100$ Gbps user experienced data rate will be required, respectively.
\vspace{1mm}

\begin{figure}[t]
    \centering
    \includegraphics[width=1\columnwidth]{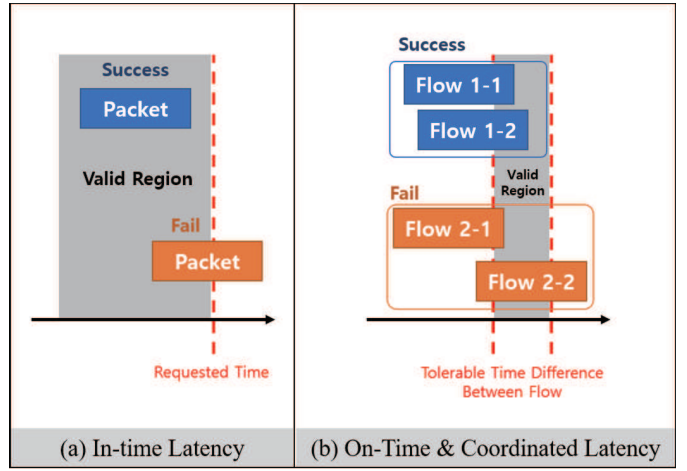}
    \caption{Example of On-time and Coordinated Latency.}
    \label{fig_4}
\end{figure}

\begin{itemize}[leftmargin=*]
\item \textbf{Insensible On-time and Coordinated Latency}
\end{itemize}

Before 6G, real-time services led the main stream in wireless communications, which focused on improving in-time latency. However, in 6G, new kinds of applications including hologram and tactile internet strongly will require on-time latency and coordinated latency which means multiple data flows arrives at a specific time with small time difference as shown in Figure 1 \cite{ITU}. To be specific, multi-sense hologram services must be synchronized in terms of visual, smell, touch information. This is a very tight requirement to consider the relationship between data flows, that guarantees an insensible latency below $100$ micro seconds, that is completely unrecognizable to humans as $1/10$ of the 5G target in-time latency.

\begin{itemize}[leftmargin=*]
\item \textbf{Scalable Connectivity}
\end{itemize}

With the explosive increase in machine-made-traffic, machine-to-machine communications will lead to the expansion of small-scale networks such as body networks, vehicle networks, and connected-machine networks. While the 5G IoT focuses on massive connectivity for small packets, 6G machine-made traffics will require high data rates at the same time as well as connectivity. In order to satisfy this harsh requirement, 
instead of centralized network management, an event-driven ecosystem in which the machine itself can freely connect and disconnect should be formed. Hundreds of devices will organize a small cluster network by themselves, and millions of clusters will be connected to each other. Finally, more than $10^8$ terminals will be directly or indirectly connected. 

\begin{itemize}[leftmargin=*]
\item \textbf{Enhanced Coverage \& Mobility}
\end{itemize}

With the diversification of transportation, the demand for aviation and maritime internet is exploding, and there is a need for a universal access system that can support Internet connectivity even in areas without wired infrastructures. This is not limited to the ground but will need to be expanded to three-dimensional (3D) coverage including the space, air, and maritime realms. As the unmanned vehicle increases, mobility should be viewed as an opportunity to form a new network architecture, not just to overcome.

\begin{figure}[t]
    \centering
    \includegraphics[width=1\columnwidth]{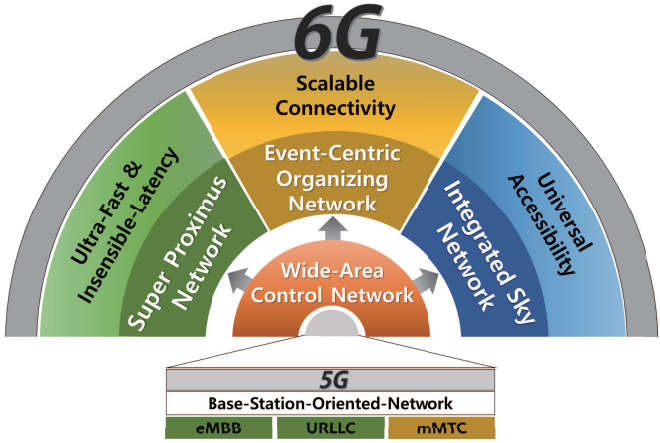}
    \caption{Network Diversification: 6G network is diversified according to target requirements.}
    \label{fig_1}
\end{figure}

\section{Network Diversification}
The extreme data rate, flexible connections of intelligent machines, and universal accessibility required in 6G will be very difficult to achieve with the current BS-oriented network architecture. For example, supporting ultra-fast services requires providing ultra-high-quality links. However, since mobility is also a very important factor in current cellular communication systems, focusing only on improving the extreme data rate is a difficult proposition. Moreover, in a situation where millions of devices transmit and receive big data simultaneously, it is challenging for a BS to support all of them in a very short time window. In terms of coverage, cellular systems are limited when it comes to unfavored and isolated regions with little or no communication infrastructure. Even though the diversity of services is increasing, the diversity of network structure is not ready to support. In this regard, 6G systems will need diversified structures that consist of specialized networks for specific target scenarios. 

Along these lines, we propose a new network architecture consisting of a super proximus network (SPN)\footnote{\textit{Proximus} means spatial and temporal closeness in a Latin word.}, an event-centric organizing network (EON), an integrated sky network (ISN), and a wide-area control network (WCN). As shown in Figure 2 and 4-(a), the proposed networks are specialized to meet specific and extreme requirements: i) ultra-fast and insensible-latency transmission, ii) scalable connectivity, iii) universal accessibility, and iv) network management. The four types of network architectures are elaborated according to target services, communication range, and link characteristics as follows:

\subsection{Super Proximus Network}

The emergence of new holographic media and massive connected machines will require a more than thousand-fold (1000x) increase in network capacity over 5G. Increasing network capacity dramatically will require comprehensive improvement in the three areas of communication distance, signal quality, and number of links. A SPN composed of programmable radio walls called smart surface, which is a specialized approach for achieving the extreme high data rate needed to support human activities in the both indoor and outdoor environments~\cite{Surface}.

SPN provides a Tera Hertz (THz) friendly environment with rich LOS as well as a short distance between the terminal and the radio interface. In addition, as shown in Figure~3, the intended multi-path can be generated through cooperative transmission of multiple smart surfaces to fill coverage holes. This serves as a medium for transmitting strong radio waves via multiple channels as well as for addressing the physical limitations of high-frequency radio stations with short coverage~\cite{Surface}. In addition, it can be used as an interface for wireless energy transfer by taking advantage of the proximity between surface interfaces and users.

The SPN includes an ultra-massive multiple-input-multiple-output (UM-MIMO) element, which can reliably support a semi-dedicated link by sharp beam-forming. By tracking the location of the terminal, it is possible to provide a stable and personalized link to vastly expand the network capacity. It also acts as a common interface that can accommodate not only the 3rd Generation Partnership Project (3GPP) but also Non-3GPP devices supporting unlicensed bands.

Figure 4-(b) shows the key performance indicators for SPN, which specializes in providing ultra-fast data transmission in low-mobility environments. With a high modulation technique of over $1024$ QAM, $64$ MIMO layer, and $10$ GHz bandwidth, SPN provides a peak data rate of $10$ Tbps. Thanks to the fact that there are plenty of THz bands, a user experienced data rate of $1$ Tbps can be achieved by allocating $1$ GHz bandwidth per user. In addition, by providing a very short transmission time interval (TTI) through ultra-short-range communication between users and the surface interface, it is possible to provide 0.01-millisecond insensible latency services with extremely fast responses.

\begin{figure}[t]
    \centering
    \includegraphics[width=1\columnwidth]{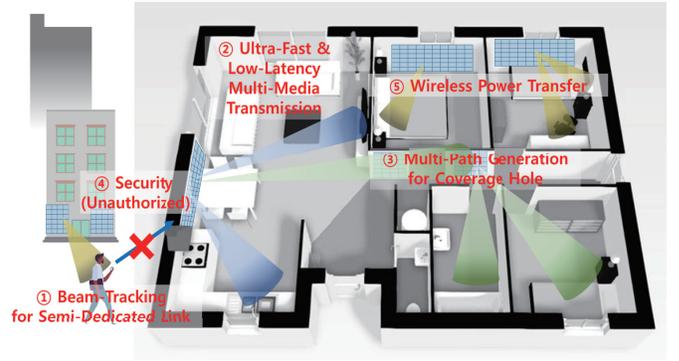}
    \caption{Use cases of Smart Surface in SPN.}
    \label{fig_2}
\end{figure}

\begin{figure*}[!t]
    \centering
    \includegraphics[width=2\columnwidth]{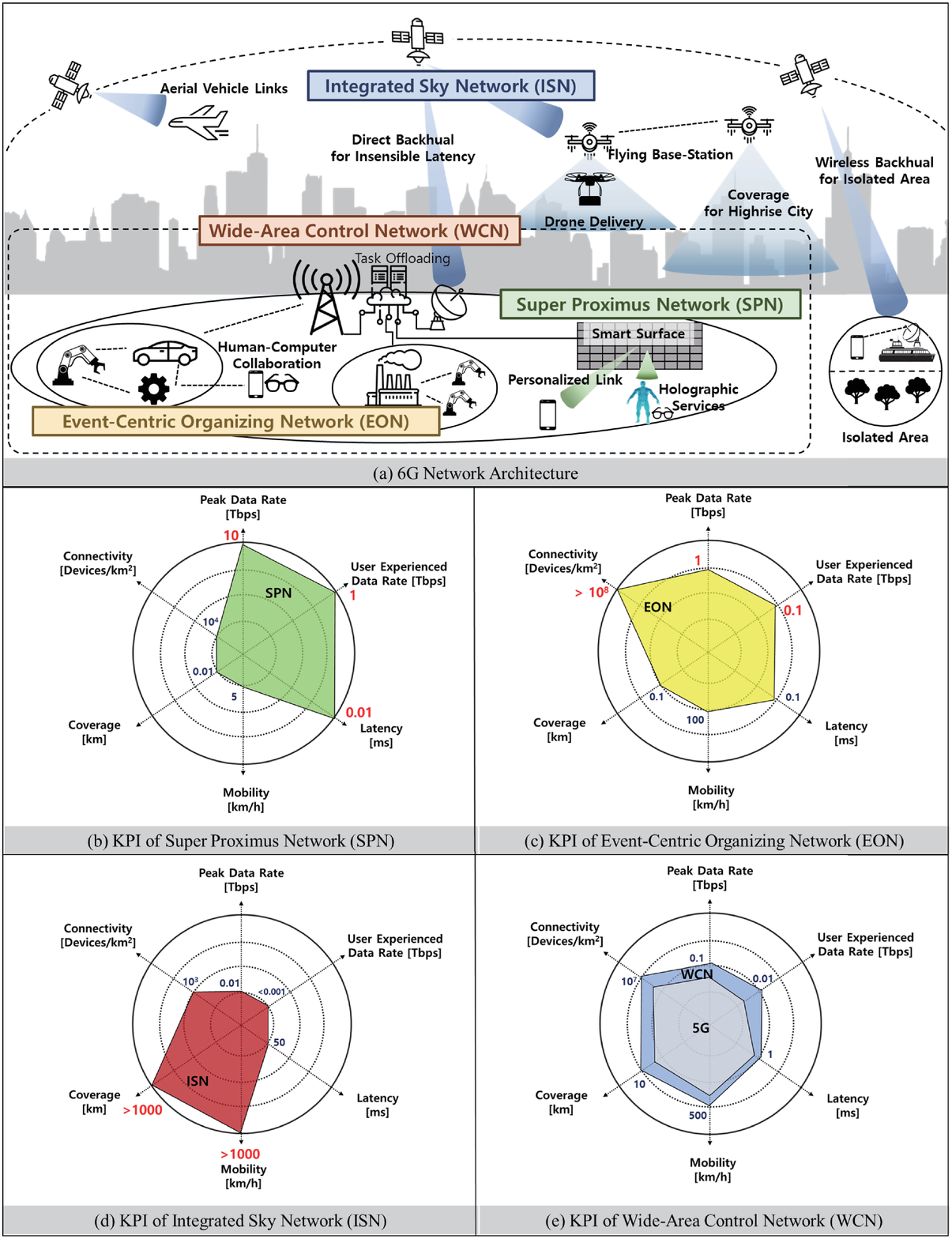}
    \caption{6G Network Architecture and Key Capabilities.}
    \label{fig_3}
\end{figure*}

\subsection{Event-Centric Organizing Network}

In the 5G mMTC scenario, since IoT devices transmitting small size packets were the main service target, it was possible to control massive links via the cellular network. But with the growing number of machines actively generating and distributing data, a centralized cellular network would have a very difficult time serving and controlling millions of devices. Also, in terms of network scalability, the centralized approach is not suitable for the HCC and RPA being considered in 6G. In this flow, 6G requires an event-centric organizing system where devices form a localized network by themselves whenever any events request communications. In EON, the target of the connection includes infrastructures as well as machines. Through connections with infrastructures, machines can also communicate with other types of networks such as SPN, WCN, and ISN.

EON has a hierarchical structure, in which devices form a cluster, share and process data within the cluster, and pass them on to other networks. It has the feature of being able to establish or disconnect links flexibly like an ecosystem. EON supports omnidirectional communications by supporting multiple direct links. Since this is not a cellular system in which the roles of BS and user are fixed, the concept of uplink (UL) / downlink (DL) is no longer meaningful. In addition, via block-chain-based link connection, EON can flexibly expand while maintaining strong security.

EON combined with satellite systems will be also very effective in drone delivery and next-generation vehicle communication systems without coverage holes. In addition, given the explosive increase in demand for private networks, including smart factories, it is worth paying attention to EON as a cost-efficient solution for organizing a localized network.

Figure 4-(c) illustrates the key capabilities of EON, which specializes in providing scalable connectivity for HCC and RPA systems. Devices within $100$ meters of each other are directly or indirectly connected in the form of clusters or platoons, and millions of clusters communicate. As a result, more than $10^8$ devices are connected per square kilometer. Although the coverage of a cluster is only $100$ meters, it has infinite scalability via interconnections among other clusters and infrastructures. In addition, multiple direct links make it possible to provide $1$ Tbps peak data rate with a low latency of $0.1$ milliseconds.

\subsection{Integrated Sky Network}

In 6G, users will not only demand high-quality communication on the ground but also in maritime, aviation, desert, and mountainous areas. This requirement is difficult to meet with the existing terrestrial networks because they require a fixed infrastructure including fiber cable, gateways and BSs. Installing such devices in isolated areas takes an inordinately long time and is extremely costly. To overcome this constraint, ISN is needed. It provides universal accessibility using geo-synchronous Earth orbit (GEO) satellites, medium Earth orbit (MEO) satellites, low-Earth orbit (LEO) satellites, and unmanned aerial vehicles as flying BSs. This network will be useful for connecting communications services in developing countries and will also help support emergency communication in severe disasters.

\begin{table*}[t!]
    \centering
    \caption{Summary of 6G networks.}
    \includegraphics[width=2\columnwidth]{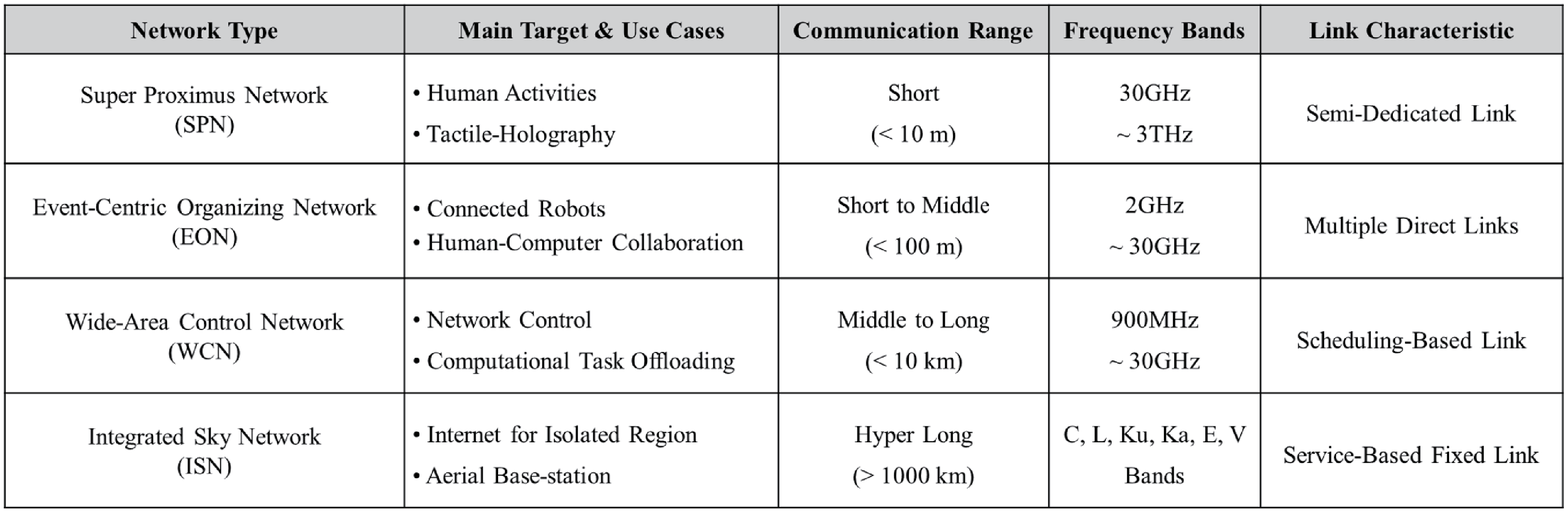}
    \label{t_1}
\end{table*}

The integrated sky network has the additional advantage of guaranteeing low latency. In the conventional terrestrial network, signals pass through a number of network nodes, causing inevitable nodal delays. Moreover, signals travel through fiber optic cable as a microwave. They propagate at only two-thirds the speed of light due to the resistance and refraction in the cable. In addition, wired interfaces depend on their network topology for latency performance. In contrast, ISN enables end-to-end communication through relays between flying BSs. The medium between sky stations and terminals is air. Therefore, signals can be transmitted without speed loss in the network. This ensures a significantly low response time anywhere in the world. In ISN, unlike ground networks, since the stations making up the ISN continue to move, cooperation and handover technologies among satellites/flying BSs will need to be applied so as to support stable and reliable services to users.

Figure 4-(d) shows the key performance indicators for ISN, which specializes in universal accessibility and high-mobility applications. Through its global multi-satellite coverage with regional beamforming technologies, ISN can provide seamless network services across distances of thousands of kilometers. It also provides stable data service for high speed objects moving at more than $1000$ km/h by utilizing position-based beam tracking technology. In addition, ISN serves as a direct backhaul with a latency performance of $50$ milliseconds, providing delay-sensitive services through cooperation with mobile edge computing (MEC) systems.

\subsection{Wide-Area Control Network}

The architecture of the WCN for 6G is similar to the existing 5G cellular network. The main role of WCN as distinguished from the 5G cellular network is to control the three networks mentioned above. Since it is difficult for distributed EON to perform network orchestration on its own, WCN serves as a control unit which is responsible for coordination among EONs. Moreover, if a device communicating with other devices via EON moves to an isolated area, it may need to change its connection to the ISN via satellite communication. We call this inter-network management a network transition. During the network transition, it is also the role of WCN to manage which network it should connect to, depending on the location of the device, the environment, and the services required.

Another role of WCN is reducing the burden on other networks by computing on behalf of other networks. Localized networks require AI-based computing, caching, and control functions that can handle data on their own. However, if the localized network is not able to handle the data (e.g., because of a lack of computing power), this can be done in the WCN instead.

The last role of WCN is to cover a wide area and support the links that cannot be connected to the other three networks. For example, a device with medium-level mobility such as a vehicle cannot satisfy the communication quality required by SPN or EON alone. In addition, WCN ensures a stable connection over a wide area, bridging the gap during the network transition so that the connection is not lost.

Figure 4-(e) depicts the key capabilities of WCN, which basically performs the existing functions of the cellular network and is in charge of controlling the network. With the development of advanced MIMO, coding schemes, flexible frame structure, and AI technology, it will achieve a ten-fold performance improvement over the 5G system, thereby covering the wide communication area that SPN and ISN cannot service. As an additional function, it is responsible for controlling the entire network and providing processing power.

\section{Research Issues and Challenges on 6G}
Of the requirements outlined in Sections I and II, what naturally gets the most attention is the need for radically higher data rates across the board and seamless connectivity. Our view is that this requirement can be met through combined gains in three categories: air technologies, network technologies, and assistive technologies.

\subsection{Key Technologies for SPN}

\begin{itemize}[leftmargin=*]
\item \textbf{Air: THz Transmission}
\end{itemize}

\begin{table}[t]
    \centering
    \caption{FCC FREQUENCY ALLOCATIONS over $30$GHz.}
    \includegraphics[width=1\columnwidth]{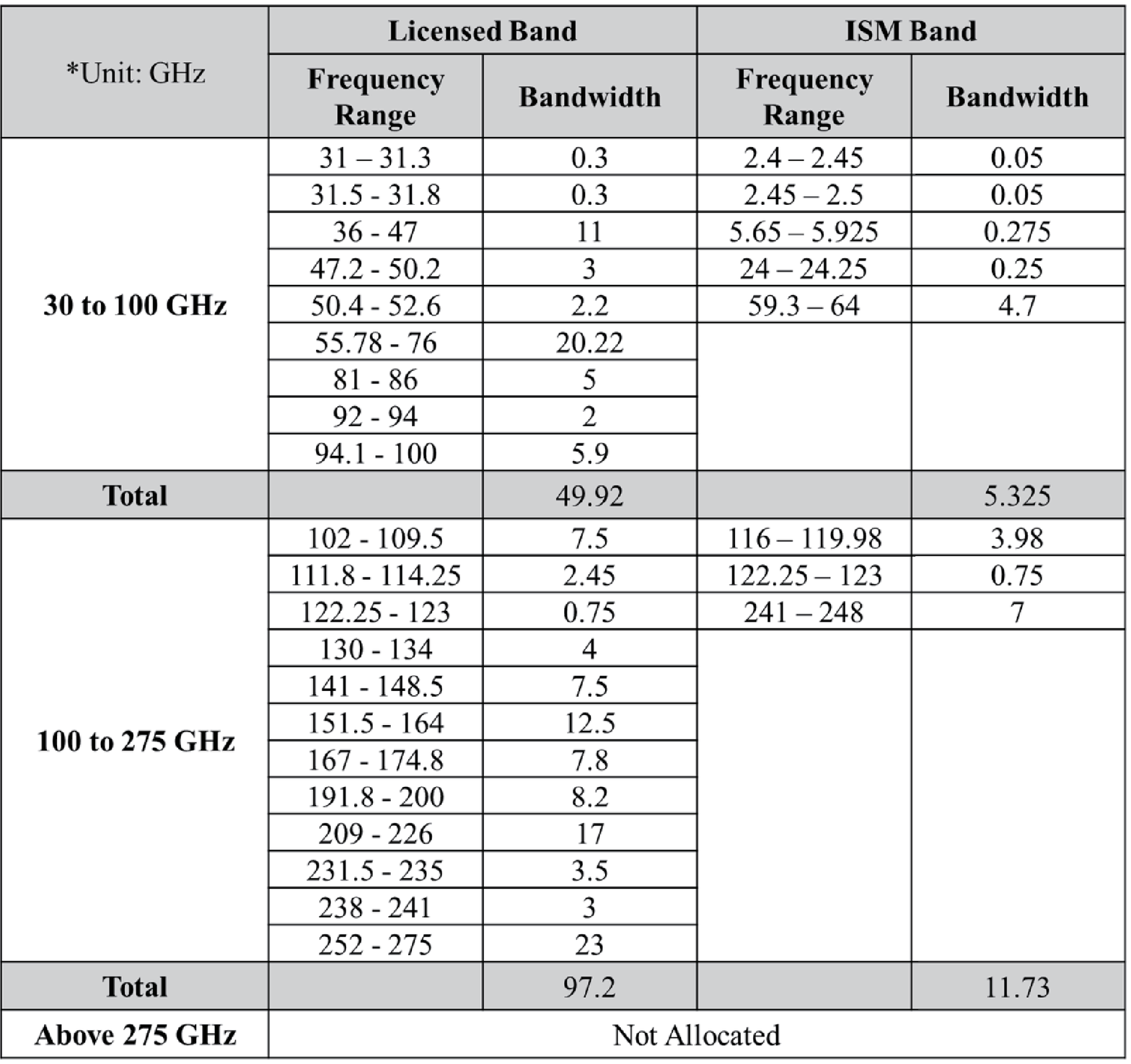}
    \label{t_2}
\end{table}

In order to achieve ultra-fast transmission and ultra-massive connectivity for SPN and EON, discovering new bands must take precedence over everything else. In 2018, The U.S. Federal Communications Commission (FCC) initiated regulatory discussion on expanding the spectrum above $100$ GHz which is unexplored and under-developed for mobile communication systems~\cite{THz}. Recently, as shown in Table~II, large contiguous frequency blocks have been allocated for mobile communications, including 97.2 GHz for licensed applications and 11.73 GHz for the Industrial, Scientific, and Medical (ISM) uses~\cite{FCC}.

The main obstacles in THz transmissions are high free space loss, significant diffuse scattering, frequency-selective molecular absorption, and physical blockage. While it is not easy to overcome all the challenges, one plausible solution is to create a complementary protocol for cooperation with SPN at the THz and WCN at the microwave level~\cite{THz}. In addition, SPN with UM-MIMO can be considered a practical approach for combating high propagation loss and blockage problems. Furthermore, THz transmission consumes a lot of energy for wide-range signal processing, so energy efficient processing technology such as a few-bit analog-digital converter, low-complexity UM-MIMO algorithms, position-based beam tracking, and nanostructure transceivers must be applied.

\begin{itemize}[leftmargin=*]
\item \textbf{Network: Shareable Architecture}
\end{itemize}

Small cell-related technologies have been actively developed in 4G and 5G. However, the major obstacle of deployment cost makes actively adopting it a difficult prospect. It is inefficient to install smart surfaces for each operator in a limited surface space. This problem is best resolved by designing the smart surface as a common entity. The smart surface has a common access point to manage resources and links in an integrated way. Except for the policy control and user plane functions that each operator needs independently, the control functions are handled in the common control plane. In addition, a security function will need to be adopted between the core networks and the common control plane to guarantee security between operators.

\begin{figure*}[t!]
    \centering
    \includegraphics[width=2\columnwidth]{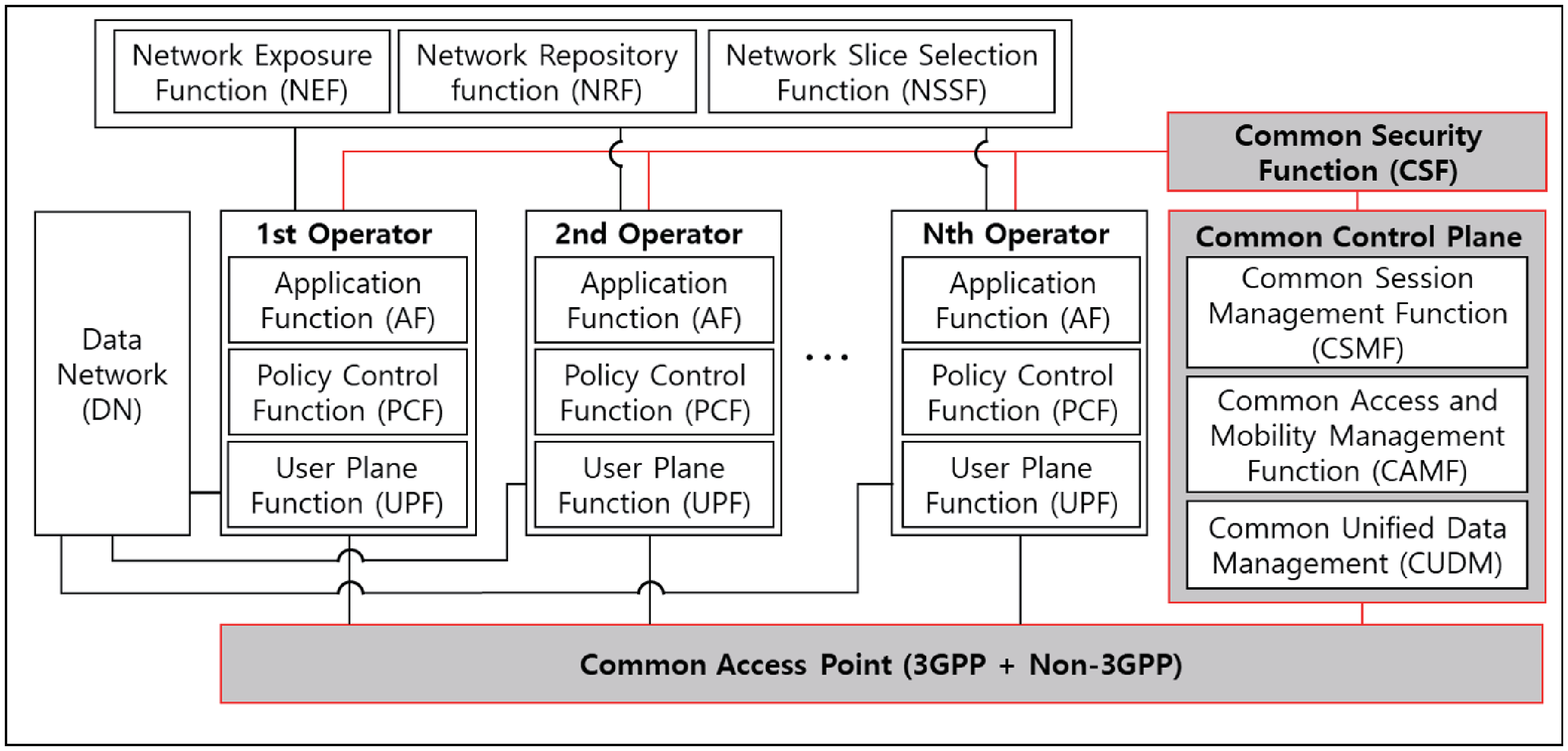}
    \caption{Simplified Example of Shared Architecture.}
    \label{fig_5}
\end{figure*}

\begin{itemize}[leftmargin=*]
\item \textbf{Assistive: Wireless Energy Transfer and Harvesting}
\end{itemize}

With the ongoing worldwide development of IoT, an unprecedented number of IoT devices will be consuming a substantially greater amount of energy~ \cite{Energy}. Along with energy transfer, energy harvesting technology, which charges batteries using ambient energy, has also attracted attention. Harvesting and sharing energy is essential for sustainable systems like IoT, which are energy-critical.

A key concern for wireless charging is the attenuation of energy being transferred over a long transmission distance. A radical approach to enhancing wireless charging efficiency is to deploy multiple antennas at the transmitter and receiver, resulting in a MIMO system. In SPN in particular, path-loss attenuation would be marginal thanks to the ubiquitous deployment of smart surfaces with UM-MIMO, which will in turn improve the charging efficiency.

\subsection{Key Technologies for EON}

\begin{itemize}[leftmargin=*]
\item \textbf{Air: Agile Waveform}
\end{itemize}

Many of the proposed new waveforms still share a common fundamental principle with OFDM~\cite{Waveform, Waveform_2}. By extension, this approach also retains the following limitations inherent in conventional multi-carrier systems: i) increasing Fast Fourier Transform size exponentially increases computation complexity, ii) although multiple numerologies have been studied, it is difficult to create different resource grids to meet different requirements simultaneously and flexibly. 

To meet this demand, a novel \textit{Agile} waveform that is flexible and adaptable to the actual network requirements will need to emerge. Especially for ultra-wide-band networks, a waveform should be designed so as to be easily detectable from the partial observation of broadband signals. The waveform will need to be flexibly and adaptively designed to support a \textit{Tetris-like} flexible resource grid to satisfy the needs of each terminal. Naturally, protocols and controlling methods should be studied to ensure that flexible resource configurations operate efficiently.

\begin{itemize}[leftmargin=*]
\item \textbf{Air: Asynchronous Transmission}
\end{itemize}

As the types of services vary, users will employ various symbol lengths to meet their latency requirements. For example, in networks requiring very high connectivity like EON, a grant-free system will be strongly considered. It is very difficult to achieve time/frequency synchronization. Coping with these difficulties will require a good deal of advancement in asynchronous transmission technology.

A challenging issue is the need to develop an asynchronous transmission protocol that is robust to various cases of interference patterns with considering an agile waveform jointly. Moreover, due to the autonomous transmission by the users, the user activity must be detected before multi-user signal decoding~\cite{Asynch}.

\begin{itemize}[leftmargin=*]
\item \textbf{Network: Block Chain-based Security}
\end{itemize}

In 6G networks, EON is particularly vulnerable to security because it has the characteristic of forming a network freely without centralized control. Decentralized coordination could be a promising solution to support security issues for scalable networks. A block-chain technology, which exploits smart transactions based on a decentralized consensus mechanism, may be worthy of attention~\cite{Security}. The block-chain can retain storage without a trusted third party and provide solid security via a tamper-resistant mechanism. However, designing a protocol that is compatible with heterogeneous types of networks is a challenging issue.

\begin{table*}[t!]
    \centering
    \caption{Summary of Key Technologies.}
    \includegraphics[width=2\columnwidth]{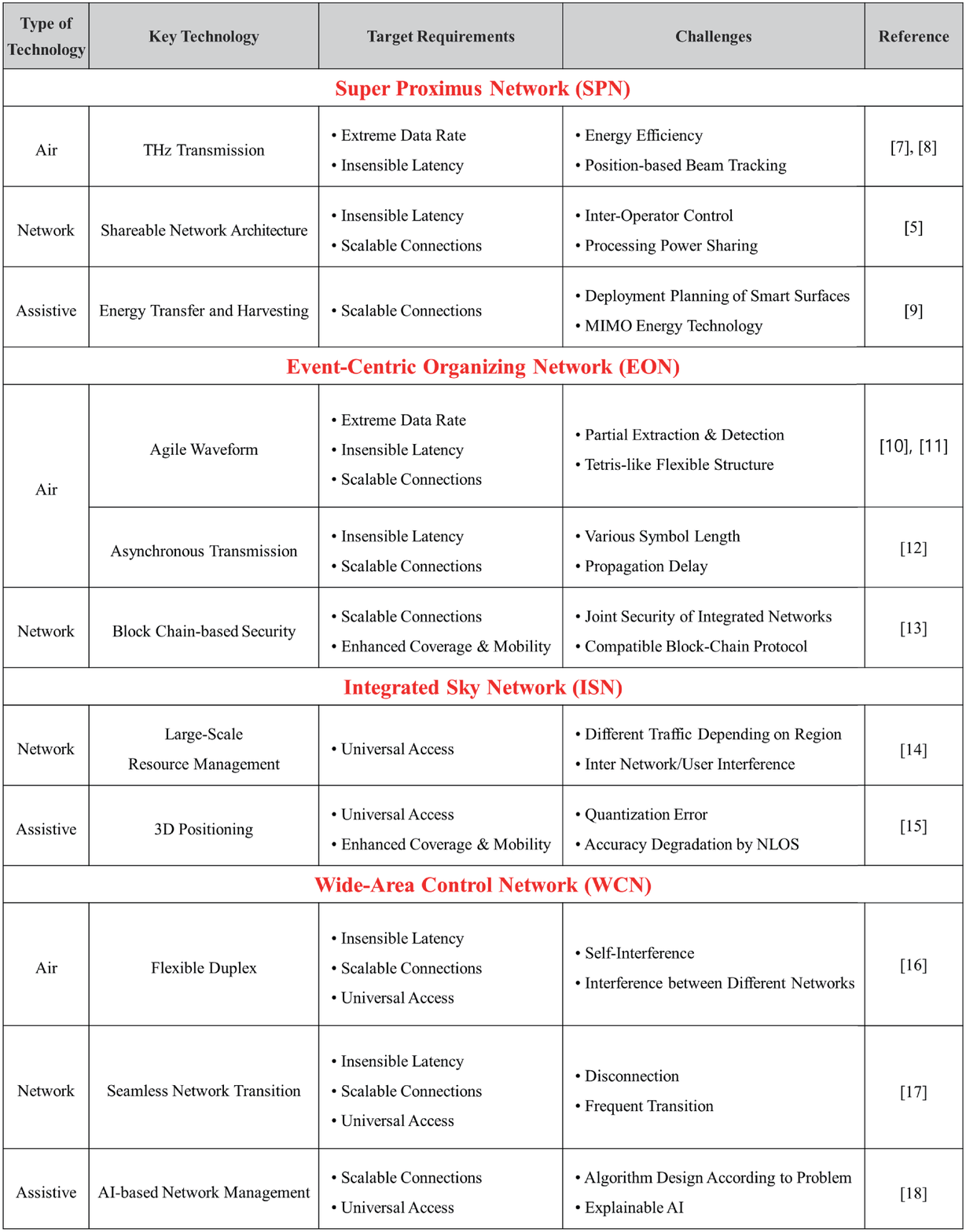}
    \label{t_3}
\end{table*}

\subsection{Key Technologies for ISN}

\begin{itemize}[leftmargin=*]
\item \textbf{Network: Large-scale Resource Management}
\end{itemize}

Resource allocation can be divided into two types: small-scale and large-scale. Small-scale resource allocation involves managing resources so as to satisfy the QoS of users in a network. Resources are allocated to users based on their requirements, such as latency, data rate, and reliability. Large-scale resource allocation involves distributing resources among networks. For the allocation to be effective, classifying and predicting network traffic is highly important. Based on that analysis, resources are allocated to meet the purpose of each network. In MEC systems, dynamic resource allocation can facilitate computation offloading, thus reducing the latency and computational burden on user terminals. Moreover, it also increases resource utilization by allowing major network operators to share resources. Such dynamic resource allocation can be made possible through network virtualization and network slicing.

\begin{itemize}[leftmargin=*]
\item \textbf{Assistive: Fine 3D Positioning}
\end{itemize}

As intelligent machines perform complex tasks in 6G, there will be a need for a way to precisely control and communicate with them. In particular, compared to previous communication systems, the target devices will not be limited to ground-based units, but rather will include drones and flying BSs as well. As a result, 3D positioning technology will be the focus of a great deal of attention for the 6G network. Furthermore, location information can be applied to precise beam targeting technologies and wireless energy transfer technologies.

In outdoor environments, according to United States Global Positioning System (GPS) data, the global average user range error is about $4.9$ meters due to signal blockage from things such as buildings, bridges, and trees. Furthermore, multi-floor 3D location is more difficult in indoor environments where GPS is not available. In order to overcome these physical limitations, compensation techniques using various types of sensor information, including ultrasonic, radar, and gyro sensor, have been studied~\cite{Positioning}.

\subsection{Key Technologies for WCN}

\begin{itemize}[leftmargin=*]
\item \textbf{Air: Flexible Duplex}
\end{itemize}

In 5G, dynamic time division duplex is supported as a way to reduce latency and more efficiently utilize spectrum by reflecting traffic demands. In 6G, however, we will need to go beyond UL/DL in cellular networks so as to coordinate duplex operation between many kinds of links~\cite{Duplex}. For EON, where machines are intricately connected in what most resembles a neural-inspired structure, multiple direct connections assume an importance that goes beyond the concept of up/down directions. Moreover, SPN, which provides interactive services with ultra-low-latency, requires duplex technology along the lines of full duplex (FD). In flexible duplex system, it will play a key role in improving data rate and reducing latency by flexibly operating on traffic and services without using a fixed duplex configuration.

\begin{itemize}[leftmargin=*]
\item \textbf{Network: Seamless Network Transition}
\end{itemize}

In 6G, where several networks with different characteristics coexist, network transitions, i.e., handovers from one type of network to another, occur according to the link environments and target services. There is currently a need for a technology that enables seamless network transition without disconnection. This ability to make seamless network transitions can improve the link coverage without limiting it to a specific network, and orchestration can be supported across different networks.

Nevertheless, keeping the number of transitions as infrequent as possible should be a priority. Having frequent transitions requires a great many control signals and increases the chance of disconnection. The network transition should be realized by ensuring that the device is connected not just to only one network, but rather is associated with several networks. As a promising solution, by applying network function virtualization (NFV), network orchestration can be realized through seamless network transition.

\begin{itemize}[leftmargin=*]
\item \textbf{Assistive: AI Techniques}
\end{itemize}

In 6G environments where network connectivity is complex and diversified, attempts are underway to integrate AI and machine learning technologies such as deep learning neural networks and reinforcement learning algorithms to manage networks more efficiently. In machine learning, the optimal solution needed to control the networks can be obtained via neural network learning instead of complex calculation. These machine-learning-based methods are the best candidates for improving the design and optimization of wireless communication systems. In particular, the main problems of synchronization, channel estimation, equalization, MIMO signal detection, and multi-user detection in wireless communication systems are being actively researched from the machine learning perspective~ \cite{AI}.

However, even though the current generation of AI systems offers tremendous benefits, their effectiveness is limited because they cannot explain the reasoning behind their decisions. To solve this problem, explainable AI has been widely studied as a way to understand the context and environments in which machines build an explanatory model that can characterize the real world. This technology helps to quickly identify the cause of network problems and has the advantage of allowing AI to communicate and collaborate with other networks for network management.

\section{Conclusion}
These are currently very exciting and important times when it comes to sketching out the concept of the future 6G. As creative and innovative convergence technologies continue to sprout up, much discussion will be needed to assess how the 6G system should change to meld them together. As this article has highlighted, the need for Network Diversification, where various types of networks operate in complement to each other and operate as a single integrated network, is key. Many technical challenges remain for implementing 6G networks in the areas of air, network, and assistive technologies. We hope this article will help serve as a basis for exploring unique innovations and moving toward a road map for the 6G network of the future.


\begin{IEEEbiography}{Hyunsoo Kim} (S'12)
received the B.S. degrees in the School of Electrical and Electronic Engineering from Yonsei University, Seoul, Korea, in 2012. He also received the Scholarship of National Research Foundation of Korea during his B.S studies. He is working toward the Ph.D. degree in Electrical Engineering at Yonsei University. His research interests are new waveform, non-orthogonal multiple access, licensed assisted access, and 6G wireless communications.
\end{IEEEbiography}

\begin{IEEEbiography}{Taehyung Kim} (S'15)
received the B.S. degree in Electrical and Electronic Engineering from Yonsei University, Seoul, Korea, in 2015. He is working toward a Ph.D. degree in Electrical and Electronic Engineering at Yonsei University, Seoul, Korea. 
His research interests are high speed train (HST) system, machine learning, cellular-based V2X, new waveform design, and non-orthogonal multiple access (NOMA).
\end{IEEEbiography}

\begin{IEEEbiography}{Hyejin Kim} (S'15)
received the B.S. degree in Electrical and Electronic Engineering from Yonsei University, Seoul, Korea, in 2015. She is working toward a Ph.D. degree in Electrical and Electronic Engineering at Yonsei University, Seoul, Korea. Her research interests are in 5G and 6G wireless communications. In addition, her current research interests include inter-cell interference management and link adaptation in the dynamic TDD systems, and new waveform design such as Windowed-OFDM, Filtered-OFDM, and FBMC.
\end{IEEEbiography}

\begin{IEEEbiography}{Insik Jung} (S'14)
received his B.S. degree in Electrical and Electronic Engineering from Yonsei University, Seoul, Korea, in 2014. He is working toward a Ph.D. degree in Electrical and Electronic Engineering at Yonsei University, Seoul, Korea. His research interests are in B5G and 6G wireless communications. In addition, his current research interests include asynchronous NOMA, FBMC, artificial intelligence and mobile edge computing. 
\end{IEEEbiography}

\begin{IEEEbiography}{Hakkeon Lee} (S'15)
received the B.S. degree in electrical and electronic engineering from Yonsei University, Seoul, South Korea, in 2015, where he is currently pursuing the Ph.D. degree in electrical and electronic engineering. His research interests are 5G and 6G wireless communications. In addition, his current research interests include resource management in the multiple access edge computing and non-orthogonal multiple access. 
\end{IEEEbiography}

\begin{IEEEbiography}{Hyunmin Seo} (S'19)
received the B.S. degrees from the School of Electrical and Electronic Engineering, Yonsei university, Seoul, Korea, in 2009. He also has been working as a RF engineer in Samsung Electronics since 2011. He is working toward the Master’s degree in Electrical Engineering at Yonsei University. His research interests are NR positioning, such as beamforming-based positioning, and 3-dimensional positioning in indoor environments. 
\end{IEEEbiography}

\begin{IEEEbiography}{Daesik Hong} (F'20, S'86, M'90, SM'05)
received the B.S. and M.S. degrees in Electronics Engineering from Yonsei University, Seoul, Korea, in 1983 and 1985, respectively, and the Ph.D. degree from the School of Electronics Engineering, Purdue University, West Lafayette, IN, in 1990. He joined Yonsei University in 1991, where he is currently a Professor with the School of Electrical and Electronic Engineering. He also served the Dean of the College of Engineering in Yonsei University from 2016 to 2019. He has been serving as Chair of Samsung-Yonsei Research Center for Mobile Intelligent Terminals. He also served as a Vice-President of Research Affairs and a President of Industry-Academic Cooperation Foundation, Yonsei University, from 2010 to 2011. He also served as a Chief Executive Officer (CEO) for Yonsei Technology Holding Company in 2011, and served as a Vice-Chair of the Institute of Electronics and Information Engineering, Korea (IEIE) in 2012. Dr. Hong is a Fellow of the IEEE. He served as an editor of the IEEE Transactions on Wireless Communications from 2006 to 2011. He currently serves as an editor of the \emph{IEEE Wireless Comm. Letters}. He was appointed as the Underwood/Avison distinguished professor at Yonsei University in 2010, and received the Best Teacher Award at Yonsei University in 2006 and 2010. He was also a recipient of the Hae-Dong Outstanding Research Awards of the Korean Institute of Communications and Information Sciences (KICS) in 2006 and the Institute of Electronics Engineers of Korea (IEEK) in 2009. His current research activities are focused on future wireless communication including new waveform, non-orthogonal multiple access, full-duplex, energy harvesting, and vehicle-to-everything communication systems. More information about his research is available at http://mirinae.yonsei.ac.kr.
\end{IEEEbiography}

%\newpage


%\vfill


\end{document}